# Conceptual Design of Mars Sample Return Mission Using Solar Montgolfieres

Malaya Kumar Biswal M[1*] and Ramesh Naidu Annavarapu[2†]

[1]Graduate Researcher, [2]Associate Professor
Department of Physics, School of Physical, Chemical, and Applied Sciences
Pondicherry University, R.V. Nagar, Puducherry – 605014, India
Email: malaykumar1997@gmail.com; rameshnaidu.phy@pondiuni.edu.in

*Abstract—* Space agencies have been incessantly working to propose a sustainable architecture for human Mars mission. But, before proceeding to a giant leap and accomplishing those mission intent, it is significant to know the extent of possibility to expand terrestrial species on the vast red planet. Decades of scientific exploration and experimentation through planetary landers and rovers showed uncertain and unsatisfactory results to determine the possibility of life on Mars. Consequently, the technological limitation has impeded to perform in-situ experimentation and analysis on the surface. Therefore, we require soil samples through Mars sample return vehicles for superior analysis in our ground-based laboratories or on-orbit analysis aboard International Space Station from the aspect of planetary protection policy. Sampling analysis either in ground or orbit will report the presence of fundamental constituents to harbor life. To effectuate this intent, we have proposed a unique sample return architecture integrated with large solar Montgolfier. Here, we deploy parachute and retro propulsion thrusters to deliver sample return vehicles on to the surface and systematic ascend with the aid of solar Montgolfier to propel the MSRV out of Mars atmosphere. Subsequently, the MSRV effectuates orbital rendezvous with orbiter for the refueling process and safe return. The proposed concept is cheap, robust and simple as compared to the current state of MSR architectures ultimately minimizing the technological necessities. It also detains backup plans that were found to be nowhere presented in any of sample return strategies. Further, we have extended our discussion to comprehensively analyze entire MSR architecture with our concept for mission feasibility.

-------------------

*Graduate Researcher, Student Member AIAA, Member of Indian Science Congress Association and Mercury Exploration Assessment Group. Department of Physics, Pondicherry University, India. Contact: malaykumar1997@gmail.com and mkumar97.res@pondiuni.edu.in.

†Associate Professor, Department of Physics, Pondicherry University, R.V Nagar, Puducherry – 605 014, India, Editorial Review and Board Member in Journal of Computational and Applied Science, Member of International Society for Gestalt Theory. Contact: arameshnaidu@gmail.com and rameshnaidu.phy@pondiuni.edu.in.

## TABLE OF CONTENTS

## 1. INTRODUCTION

Mars sample return mission is one of the significant and principal steps toward the progression of Human Mars exploration and space civilization. The investigation and analysis of soil samples on Mars are limited due to the constraints in the large dimension of the scientific instruments and the inability to deliver larger payload mass to the planetary surface. Results of robotic analysis and exploration may remain unsatisfactory and erroneous due to the damage of instrumentation during interplanetary transit caused by several space environmental factors or during the EDL performance due to hostile planetary environment. Noticeable incident is uncertainty in the report of NASA's Viking lander's biological experiment during the 1970s and the erroneous report may be due to the damage of instrumentation. Hence, satisfactory results require proper analyzing procedures in our ground-based laboratories through the return of samples by executing a Mars sample return mission. To date numerous sample return strategies have been proposed and some were under study and development (for example NASA-ESA's Sample Return Mission), but none of these strategies were executed due to their expensive nature, risk of sample handling, the threat of planetary back contamination, demand for massive and advanced technology, and sustainability. But a small step was considered and initialized with the launch of NASA's

Perseverance Rover (Mars 2020) that carried a sample cache to gather geologic samples of scientific interest from various location and finally leave it at the safest site for fetching in upcoming missions.

So to extend and support this stride of sample return campaign, we have proposed an MSR architecture using solar Montgolfieres (otherwise known as Montgolfier Orbital Rendezvous MOR-MSR approach) for returning and directing samples either to our terrestrial laboratories or on-orbit laboratory. The proposed concept is novel, cheap, feasible and sustainable minimizing the risk of mission management and back contamination from the perspective of planetary protection policy. Additionally, we have included back-up plans that were not described in any of the past proposed MSR strategies and this architecture can be called as free-sample return architecture as it goes along with systematic or conventional Mars lander missions. Finally, we expect that employing this sample return architecture, we can bring back samples by no later than the 2030s.

## 2. REVIEW ON PAST MSR CAMPAIGN [1-6]

To date, more than 20 Mars Sample Return architectures were proposed and presented at various international conferences and technical meetings by scientists and mission planners. Reviewing those plans, we found that all architectures were mainly falling under three distinct categories that include direct approach, sample return with aid of orbital rendezvous, and finally sample return using in-situ propellant production (first proposed by Zubrin and stands the current NASA-ESA concept). So, in this section, we review the overview of these mission strategies.

*Direct Approach*

A heavy-lift launch vehicle will deliver Mars Sample Return Vehicle (MSRV) comprising of Mars Ascent Vehicle (MAV), Earth Re-entry Vehicle (ERV), and a sample fetching rover to the Mars surface. After performing a successful landing, the sample collecting rover will be deployed to gather samples of scientific interest in a sample cache from various targeted sites. Then, sample collection is followed by sample loading to the MAV deploying rover's robotic arm. When the launch window for the Earth opens-up, the MAV fires and takes-off from the surface to expel out of Mars atmosphere into Mars orbit from where it follows trajectory cruise to Earth. This architecture seems simple but may cost expensive due to the transportation of MAV propellant from Earth as well as the necessity to launch two launch vehicles. Where one launch for delivering the Mars Sample Retrieval Lander (MSRL) and a Rover and another launch for capturing sample from Mars orbit and return via ERV.

*Orbital Rendezvous (Approach-A)*

This architecture initiates with two different launches where one launch delivers Earth Return Vehicle (ERV) and Entry Capsule (EC) into Mars orbit, and another launch delivers MAV and a sample handling rover to the surface. Upon successful landing, the deployed rover will collect samples and load it to the MAV. Then, the MAV uplifts and blasts-off from the surface to the orbit during appropriate Earth-Mars launch window to proceed further for orbital rendezvous with ERV, sample transfer and Earth return. Finally, the samples are directed into Earth through Entry capsule.

*Orbital Rendezvous (Approach-B)*

This approach requires three launches to complete the mission: at first launch attempt, the launch vehicle delivers the MSRL and a sample cache rover (deployed rover will start to gather samples after landing); then the second launch delivers MAV with Sample Fetching Rover (SFR) to fetch the sample cache from the desired site; the third launch delivers an ERV to perform orbital rendezvous with MAV and safe return.

Comparing orbital rendezvous approach A and B, both architecture demands the potential to launch two-three launch vehicle attempts ultimately increasing the mission budget. Similarly, the demand for designing and developing separate landers and rovers for this mission poses a challenging task from the economic and feasibility aspect. Further, executing this mission plans may expand the mission time. One of the major disadvantages of this concept is if any of the mission phases fails the whole mission remain unperturbed since mission planners have not embraced any backup plans.

*In-Situ Propellant Plant Approach*

ISPP-MSR approach utilizes in-situ propellant plant for fueling MAV instead of exporting terrestrial propellant from Earth. This architecture requires a MAV, ISPP and an SFR which are supposed to deliver to the surface of Mars in a single launch attempt. Then the deployed rover starts to gather samples and parallel, the ISPP starts producing native propellant (Methane) from atmospheric $CO_2$. After months, now the MAV is loaded with native propellant and samples and gets prepared to blasts-off from the surface during the awaited launch window (opens-up for every 1.5 years). When the launch window for Earth-Mars open-up, the MAV propels and blasts-off from the surface to the orbit and then performs safe return. This architecture is found to be simple, cheap, and effective of all approaches. But the constraints are lack of testbed for demonstrating ISPP in a microgravity environment, so it may encounter issues concerning the performance and error due to the effect of space environmental factors (the circuitries may get damaged or ruptured during interplanetary transit or due to hostile Mars environment as many spacecrafts has encountered this issues) [7-9]. In addition to this, the architecture lacks backup option and hence if the ISPP fails to produce native Martian propellant, the whole sample cache will remain on the surface until the next mission arrives to retrieve it. Hence this architecture holds a drawback and entrusts the mission reliability.

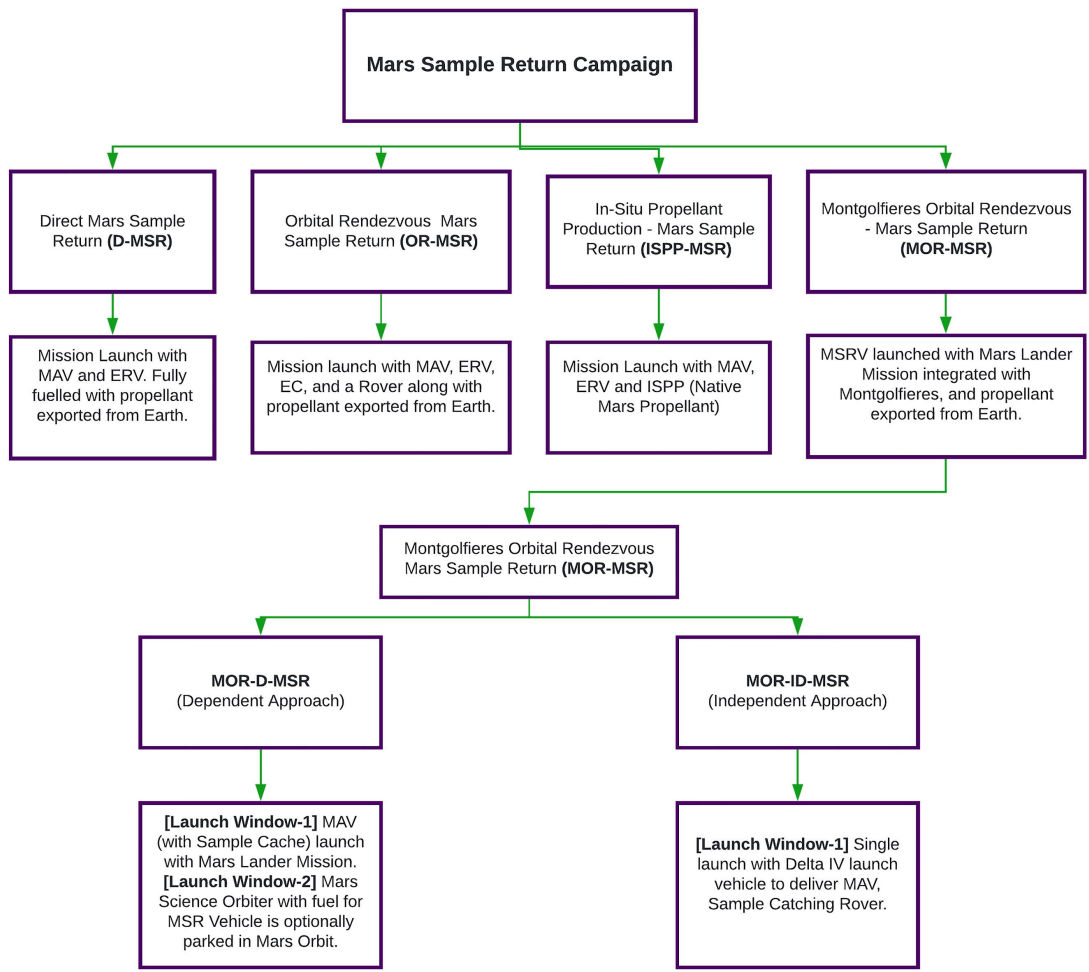

Fig 1 Outline Map for Various Mars Sample Return Strategies

Table 1 Comparative Summary of MSR Campaigns

|  | D-MSR | OR-MSR | ISPP-MSR | MOR-D-MSR | MOR-ID-MSR |
|---|---|---|---|---|---|
| No. of Launches | 2 Launches | 2-3 Launches | 1 Launch | 0-1 Launches* | 1 Launch |
| Modules Required | MAV, ERV, Rover | MAV, ERV, 2 Rover, Orbiter, Lander | MAV, ERV, Rover, ISPP | MAV, ERV | MAV, ERV, Rover |
| Ease of Complexity | Simple | Complex | Simple | Simple | Simple |
| Mission Cost | High | High | Low | Low | Affordable |
| Back-up Option | Not Available | Not Available | Not Available | Available | Not Available |
| Propellant Source | Earth | Earth | Mars | Earth | Earth |
| Proposed LV | Titan-IV | Delta-II | Delta-II | Depends | Delta-IV |

*(Optional) Deliver of Mars Science Mission Orbiter for refueling

Table 2 Technology Availability Option for MSR Campaigns

|  | D-MSR | OR-MSR | ISPP-MSR | MOR-D-MSR | MOR-ID-MSR |
|---|---|---|---|---|---|
| Separate Lander Required | Yes | Yes | Yes | No | Yes |
| Back-up Option Available | No | No | No | Yes | No |
| Orbital Rendezvous Required | No | Yes | No | Yes/No | No |
| ISS-Direct | No | No | No | Yes | Yes |
| Back Contamination Risk | Yes | Yes | Yes | No | No |
| Earth Propellant Option | Yes | Yes | No | Yes | Yes |
| Mars Propellant Option | No | No | Yes | No | No |
| Mars Orbiter Required | No | Yes | No | Optional | No |
| EEV Required | Yes | Yes | Yes | No | No |
| Rover Required | Yes | Yes | Yes | No | Yes |

## 3. CURRENT MSR CAMPAIGN

*NASA-ESA Concept*

NASA-ESA Concept intends to complete the sample return mission in three phases. The first phase is to collect samples deploying a rover called Mars 2020 to the surface. Then, NASA plans to land a Sample Retrieval Lander (SRL) to fetch sample cache (left by Mars Perseverance Rover) and deliver to Mars orbit through MAV as part of the second phase of the mission. Finally, the third phase is to bring samples back to Earth by effectuating orbital rendezvous with Earth Return Orbiter (ERO) parked in Mars orbit [10]

*Mars-Grunt (Russia)*

Mars-Grunt is the follow-up mission of Fobos-Grunt mission that lost with launch vehicle malfunction. It is a proposed concept intended to deliver ERV, MSRL, and a MAV. At the initial stage of the mission, a lander will perform soft-landing on Mars and will gather samples using its robotic arm. Sample acquisition is followed by Mars ascent where a small rocket aboard MSRL blasts-off for orbital rendezvous to ERV in orbit to transfer sample and for further Earth return. This concept is under study and feasibility of launch is uncertain but tentatively around 2026. Further, Chinese planetary exploration group also intend to attempt multiple launch MSR concept to return samples from Mars [11,12]

*MOR-ID-MSR Approach*

Montgolfier based independent Mars sample return approach uses a single launch to deliver Mars Sample Retrieval Lander (MSRL) and a Sample Fetching/Collecting Rover (SFR) to the surface. After landing, the SFR undergoes sample collection and loading to the MSRV. When the launch window opens up, the solar Montgolfier is allowed to inflate either using the solar resource or rapid inflating setup (using propellant). Upon attaining an inflated limit, the Montgolfier along with MSRV is unclasped from the MSRL. Afterwards, the Montgolfier starts to ascend to the upper atmosphere of Mars from where the MSRV propels and reaches Mars orbit. (In case of fuel shortage, the MSRV performs orbital rendezvous with a Mars Science Orbiter (MSO) which is optionally parked in orbit to for fueling MSRV if required). Thereafter, the MSRV follows trajectory cruise for safe return to Earth. One of the major disadvantages of this concept is the unavailability of backup plans if the Montgolfier fails to inflate and ascend the entire mission will remain unaccomplished.

*Architecture Assessment and MOR-MSR Proposition*

We can observe that the current strategy of Mars sample return endeavour follows orbital rendezvous approach (OR-MSR) otherwise known as multiple launch concept. The scenario of multiple launch and demand for the development of separate lander, ERV, and rover increases the complexity of mission and grasp two-three phases to complete. Hypothetically a mishap at any phase may lead to a complete decline of mission as no backup plan is proposed for these strategies. Similarly, all proposed strategies follow direct return to Earth with random screening process and this may be ineffective against back contamination and planetary protection policy. So, focusing on the complexity and drawbacks of past MSR missions, we propose Montgolfier based Mars sample return architecture (MOR-MSR) and it

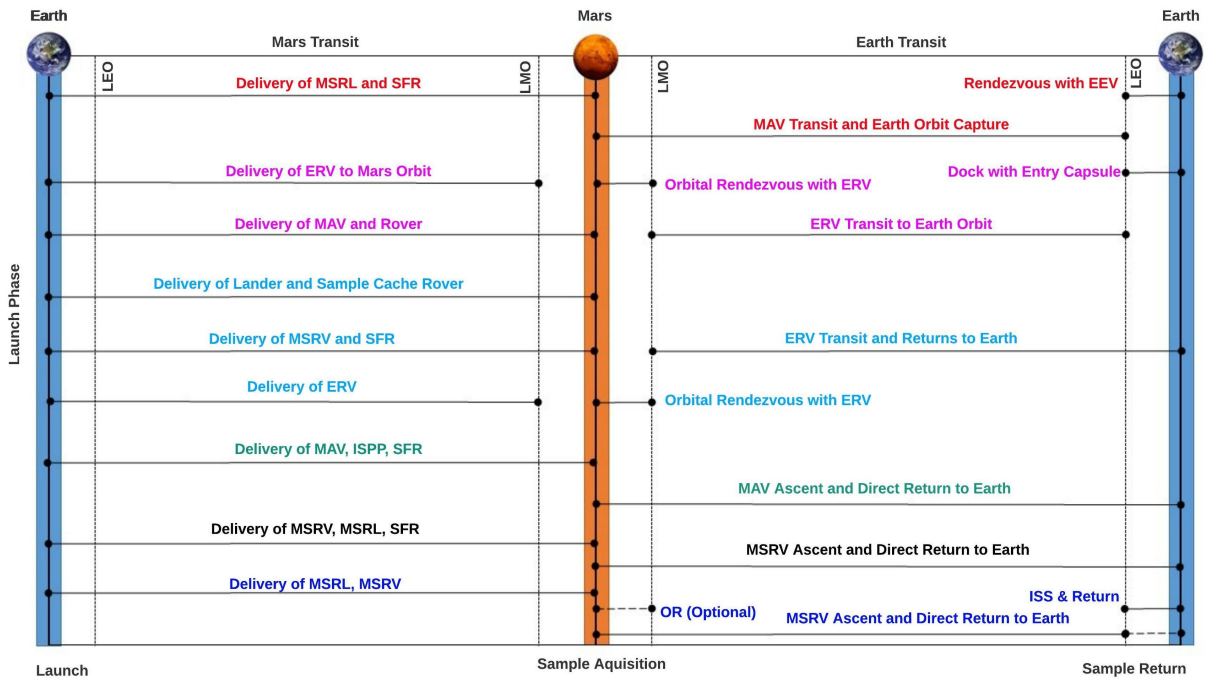

**Fig 2 Overview of Mission Sequence for Distinct MSR Strategies**

**Color Code:** ■ Direct Approach ■ OR (Approach-A) ■ OR (Approach-B) ■ ISPP ■ MOR-ID-MSR ■ MOR-D-MSR

affords a complete solution to the challenge of unavailability of backup plans, risk of back contamination, and redundant technology. Further, we have described the overview of MOR-D-MSR approach in upcoming sections. We have shown the types of various MSR campaigns in Fig 1 along with their mission timeline in Fig 2. Furthermore, from review and analysis, we have differentiated the technological requirements and availability option and in table 1 and 2.

## 4. DESIGN METHODOLOGY

*Mars Sample Retrieval Lander*

The MSR mission comprises of an MSRL along with MSRV (Mars Sample Return Vehicle). The MSRV is enclosed and integrated within MSRL. The lower deck of MSRL has four exterior propellant tank to power the MSRL thrusters to break the terminal velocity during EDL phase and four interior propellant tank strapped to the interlocking port cylinder to power MSRV thrusters as shown in Fig 3. Similarly, the lander's upper deck has space allocated for mounting science payloads and various equipment. In addition to this, the upper deck has the upper stage of MSRV which is inclined over a slider track laid along the extent of lander's diameter. The center portion of the lander has MSRV interlocking port cylinder integrated with the interior propellant tanks. Just aside the interlocking port, there is a Montgolfier stowed bag strapped to the Montgolfier strap port of MSRV. Below the stowed bag, there is an inflating setup to enlarge the Montgolfier in case of deficit solar resources. Finally, the lander is powered by large-sized deployable solar panels. The complete schematic diagram of MSRL with MSRV is shown in Fig (3-4) and (8-9), and configurational design in Fig 6.

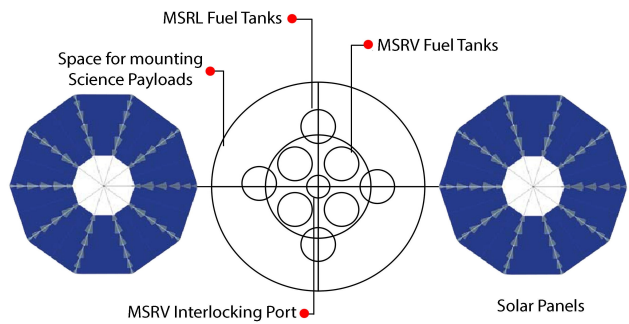

**Fig 3 Schematic View of MSRL's Lower Deck**

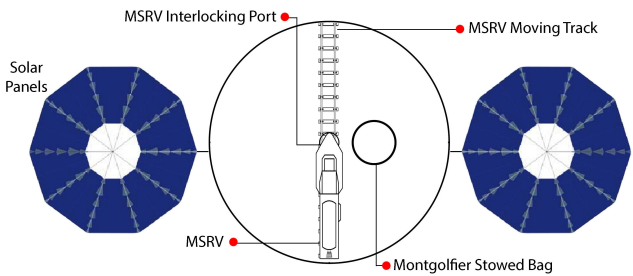

**Fig 4 Schematic View of MSRL's Upper Deck**

*Mars Sample Return Vehicle*

The MSRV has a two-stage solid or liquid cryogenic propellant engine for powered propulsion ascent. The top of the vehicle has a payload fairing cone mounted with the avionics system. The sample cache container is attached to the avionics control system and shielded with a radiation suit to avoid the damage of the vehicle subsystem due to vulnerability to interplanetary radiation [7]. It also protects the sample from radiation contamination. This configuration forms the upper stage of the Mars Sample Return Vehicle.

The upper stage is integrated with Stage-I propulsion system to power the spaceflight from Mars upper atmosphere into orbit and then to Earth (In the scenario of Montgolfier powered ascend). The stage one propulsion system along with its thrusters forms the middle stage of the MSRV. Finally, both the upper and middle stage is interlinked to the interior propellant tanks (lower stage) via interlocking port to form the complete structure of MSRV. This interlocking mechanism is supposed to execute in case of Montgolfier failure as part of the backup option. The lower stage propellant tank will power the MSRV to perform propulsive ascent from the surface to orbit. Complete configurational design of MSRV is shown in Fig 5.

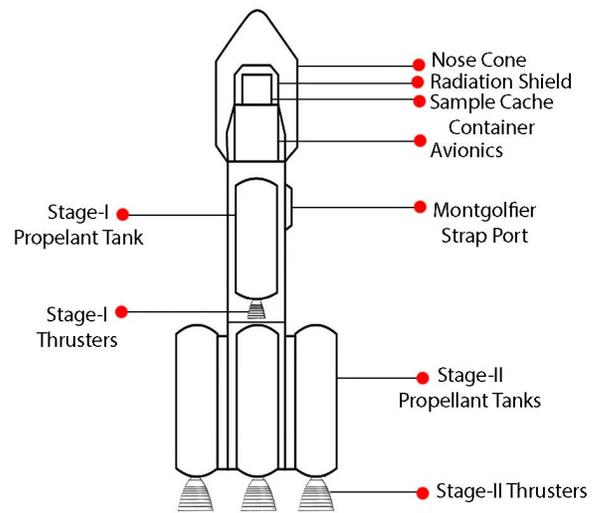

**Fig 5 Schematic View of MSRV**

*Solar Montgolfier*

The concept of Montgolfier or balloon-based planetary exploration was first put forward by Jacques Blamont and the first-ever mission was Vega Venus probe attempted during the 1980s. Later on, mission planners extended their vision of exploring Mars and other planetary atmospheres using Montgolfieres. NASA's Jet propulsion laboratory and French space agency (CNES) have validated several balloon-based experiments (for example ALICE 1993) through Mars Aerobot Validation Program (MAVP) [13] and currently, the advanced concept is being developed under MABTEX (Mars Aerobot Technology Experiment) program [14].

Hence, we intend to deploy solar Montgolfier with a volume of approximately $\geq 100{,}000$ cubic meters so that the balloon can carry up to a mass of approximately ≥250 kilograms. Further, we have provided an additional inflating setup to inflate the balloon in case of deficit solar resource to arise the balloon assembly. Furthermore, launching MSRV from balloon assembly may not be a challenging task as we have enough maturation of technology and experiences learned from past experiments. Noticeable experiments were Rockoon in 1949 and Rockoon in 2018 accomplished by Leo Aerospace LLC [15]. Artist conceptual design of planetary balloon exploration is shown in Fig 6.

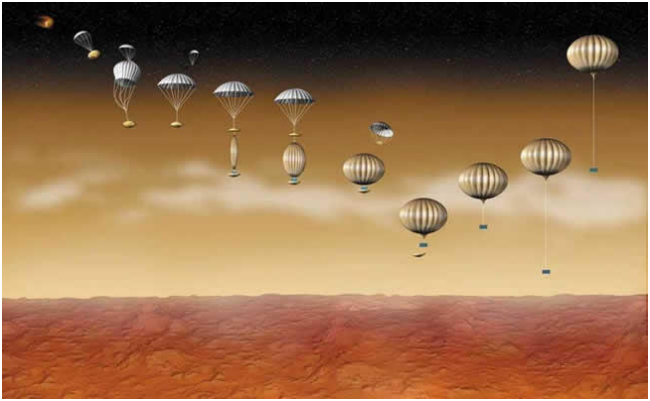

**Fig 6 Artist Concept of Planetary Balloon Exploration**

**(Image Courtesy: NASA)**

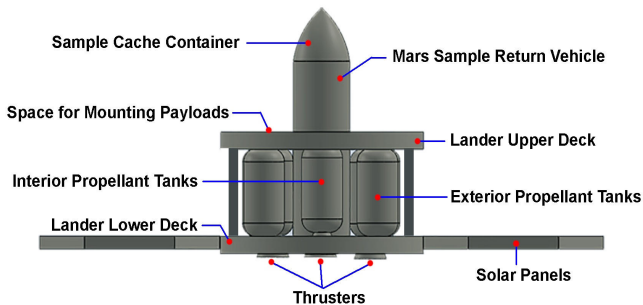

**Fig 7 Typical Design of MSRL with MSRV**

## 5. WORKING OF MARS SAMPLE RETURN VEHICLE

The MSRL upon touchdown, the sample cache will get fetched by either Curiosity/Mars 2020 or dedicated SFR and load it to the MSRV sample container. Prior to this process, the MSRV which is resting over the slider track will move aside to the top corner of lander (shown in Fig 4) to provide easy access for sample loading. Then the MSRV will open its payload fairing cone and sample container followed by sample loading via rover's robotic arm. Subsequent to successful sample acquisition, the MSRL's MSRV initializes its operation for Mars ascend.

*Scenario-I (Montgolfier Guided Ascent)*

When the launch window for the Earth transit opens-up, the MSRV get centrally aligned along with the central position and locked with the strap port. In that case, the MSRV's strap port is coupled to the Montgolfier assembly followed by balloon inflation either utilizing solar resources or the inflating setup. After attaining a certain inflation limit, the Montgolfier assembly along with its MSRV is extricated from MSRL for Mars atmospheric ascent. This event is followed by Stage-I powered propulsion and atmospheric escape from Mars into orbit.

*Scenario-II (Propulsive Ascent)*

In the case of Montgolfier malfunction, the MSRV is elevated, centrally aligned perpendicular to the lander deck shown in Fig 9, and interlocked to the interior propellant tanks through interlocking port followed by propulsive ascent with the aid of Stage-II propulsion system. After reaching a maximum altitude, the strap-on Stage-II propellant tanks get separated-off the MSRV. Mars ascent is followed by ignition of Stage-I powered propulsion for Mars atmospheric escape and orbit capture.

*Mars Orbital Rendezvous*

Before Mars ascent, the orbit of Mars is optionally parked with Mars Science Mission Orbiter (MSMO) [primarily intended for orbital observations and can be extended to refuel MSRV if required]. The MSMO is provided with additional propellant tanks to refuel MSRV in orbit (in case of MSRV fuel exhaustion) or else it can be employed to extend the mission life.

The MSRV upon achieving Mars orbit, it is allowed to perform orbital rendezvous with MSMO for on-orbit refueling. This scenario is recommended for Scenario-II (powered propulsion ascent) due to fuel exhaustion and not necessarily required for Montgolfier guided ascend. Refueling phase follows Earth departure.

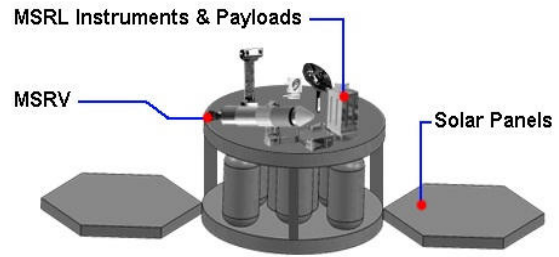

**Fig 8 Typical Design of MSRL (Before MSRV Ascent)**

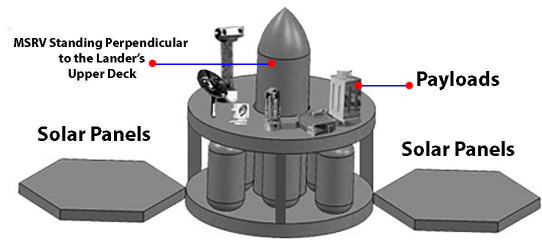

**Fig 9 Typical Design of MSRL with MSRV (Ready for Ascent)**

## 6. MISSION OVERVIEW OF MOR-D-MSR APPROACH

A heavy-lift launch vehicle (i.e. Falcon or Delta class) will deliver an MSRL together with MSRV to LEO during appropriate launch window followed by 180 days Mars Transit. Then, MSRL separates from cruise stage and performs hypersonic dive into the Mars atmosphere and achieves soft-landing thereby deploying a parachute and powered propulsion system (powered by exterior propellant tanks). Upon successful landing, the twin rovers either Curiosity/Mars 2020 approaches MSRL landing site to load the sample cache into MSRV (This sequence of events goes parallel to the objective of lander mission). When the launch window for the Earth-Mars opens-up the MSRV uplifts into the orbit through distinct ascent technique explained in Scenarios (I and II).

Subsequent to MSRV successful orbit capture, it performs orbital rendezvous with MSMO which is optionally parked at LMO or HMO in advance of MSRV ascend to refuel the vehicle (in case of propulsive ascent). Succeeding on-orbit refueling, the MSRV returns to Earth using its advanced avionics, guidance and navigation system. Upon Earth approach, the MSRV and its sample cache are directed to ISS (International Space Station) laboratory from the perspective of back-contamination where the sample undergoes critical screening process to determine the level of radiation-exposed and contaminants acquired during interplanetary spaceflight.

Consequent to the proper screening process, the sample undergoes further investigation procedures and analysis either aboard ISS under microgravity or it can be returned back to earth via Dragon capsule during crew return from ISS expedition mission (instead of using separate EC). Therefore, it reduces the cost of deploying a separate Earth Entry Vehicle (EEV) for directing samples from LEO and also minimizes the risk of back-contamination. Complete mission map is shown in Fig 10.

## 7. DISCUSSIONS

Earlier proposed concept has no backup plans as their main disadvantage and based on multiple launch concept increasing the mission cost and risk. Even the NASA-ESA concept relies on this strategy. So focusing on the complexity of past MSR campaign, our proposed concept is feasible from the perspective of current maturation of technology. Further, past strategies are associated with challenges like executing orbital rendezvous, expensive for exporting propellant from Earth. But in this decade, we have succeeded in achieving multiple orbital rendezvous and refueling experiment [8] and hence these experiences will help make our current architecture success. Therefore, our concept may increase the probability of returning samples from Mars in or around 2030s.

## 8. CONCLUSION

Mars Sample Return mission is one of the significant part of robotic as well as manned mission due to its tremendous scientific return. It enables us to decide the effectiveness of the crewed mission and aid to accomplish major objectives of the Mars Exploration Program Analysis Group (MEPAG) [16]. Focusing on the demand for high scientific return and challenges of past proposed sample return campaigns, we have proposed an architecture based on planetary balloons capable of addressing complete challenges. This concept is a derived version of direct and orbital rendezvous approach to ensure sustainability and feasibility of mission with minimal risk management. Additionally, the proposed concept is compared with other strategies to show the effectiveness of mission and technological necessities. Further, we have displayed typical design of MSRV and MSRL in appropriate section along with the detailed outline map of mission scenario shown in Fig 10 Finally, we expect that our proposed architecture may conceptualize mission planners concerning the technology required for enhancing the mission.

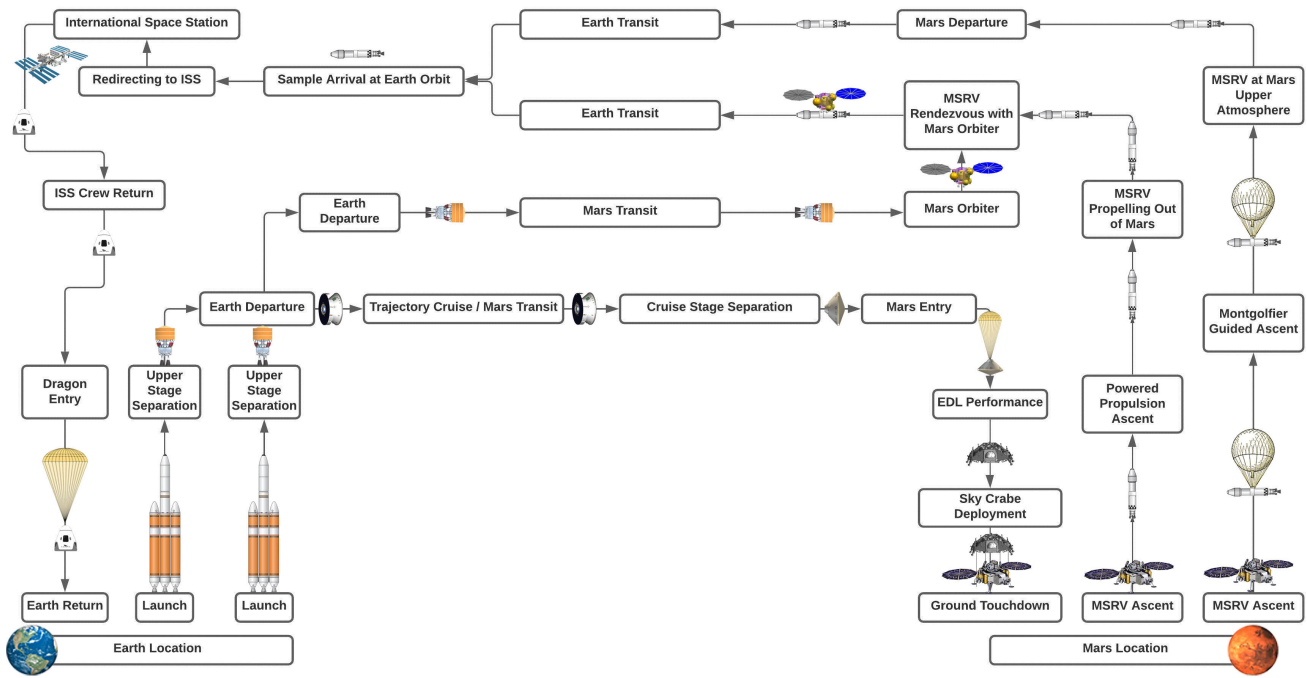

**Fig 10 Outline Map for Complete Mars Sample Return Mission Scenario**

| Event/Year | 2020 | 2020-21 | 2021 | 2022 | 2026 | 2026-27 | 2027 | 2028 | 2029 | 2029-30 | 2030 | 2030-31 | 2031 |
|---|---|---|---|---|---|---|---|---|---|---|---|---|---|
| Mars 2020 Rover Launch | ■ | | | | | | | | | | | | |
| *Mars Transit* | | ■ | | | | | | | | | | | |
| Mars 2020 Landing | | | ■ | | | | | | | | | | |
| Sample Collection | | | | ■ | | | | | | | | | |
| MSRL/MSRV Launch | | | | | ■ | | | | | | | | |
| *Mars Transit* | | | | | | ■ | | | | | | | |
| MSRL/MSRV Landing | | | | | | | ■ | | | | | | |
| Sample Deliver to MSRV | | | | | | | | ■ | | | | | |
| MSO Launch | | | | | | | | | ■ | | | | |
| *Mars Transit* | | | | | | | | | | ■ | | | |
| MSO Arrival | | | | | | | | | | | ■ | | |
| MSRV Ascent from Mars | | | | | | | | | | | ■ | | |
| *Earth Transit* | | | | | | | | | | | | ■ | |
| Sample Return to Earth | | | | | | | | | | | | | ■ |

Table 3 Mars Sample Return Mission Timeline

We have shown the estimated mission timeline for our proposed Mars Sample Return concept in table 3.

## 9. FUTURE WORK

Dimensional analysis, design and simulation methods for this concept are under study and we intend to write another paper presenting trajectory and quantitative analysis. Further, we look for opportunities to pursue this research and study in any advanced aerospace laboratories or universities. And we look forward to working together with mission experts to have a collaborative contribution to the scientific community in achieving the initial phase of human planetary exploration.

## ACKNOWLEDGEMENTS

The authors would like thank **Dr. Glen. A. Guzik. PhD (NASA)**, Propulsion Test Engineer at National Aeronautics and Space Administration for his feedback and discussion to complete this research article.

I (Malaya Kumar Biswal) would like to extend my sincere thankfulness to all of my kind and lovable friends for their financial assistance for the conference participation.

## CONFLICT OF INTEREST

The authors have no potential conflict of interest to report.

## FUNDING

No external funding was received to support this study.

## DEDICATION

I (Malaya Kumar Biswal) would like to dedicate this work to by beloved mother late **Mrs. Malathi Biswal** for her motivational speech and emotional support throughout my life.

## ACRONYMS AND SUBSCRIPTS

ALICE = Altitude Control Experiment
CNES = National Centre for Space Studies
CO = Carbon-Oxygen
D-MSR = Direct Mars Sample Return
EC = Entry Capsule
EDL = Entry, Descent and Landing
EEV = Earth Entry Vehicle
ERO = Earth Return Orbiter
ERV = Earth Return Vehicle
ESA = European Space Agency
HMO = High Mars Orbit
ISPP = In-Situ Propellant Plant
ISS = International Space Station
LEO = Low Earth Orbit
LMO = Low-Mars Orbit
LV = Launch Vehicle
MABTEX = Mars Aerobot Technology Experiment
MAV = Mars Ascent Vehicle
MAVP = Mars Aerobot Validation Program
MEPAG = Mars Exploration Program Analysis Group
MLM = Mars Lander Mission
MOR = Montgolfier Orbital Rendezvous
MOR-D = Montgolfier Orbital Rendezvous - Dependent
MOR-ID = Montgolfier Orbital Rendezvous - Independent
MSMO = Mars Science Mission Orbiter
MSO = Mars Science Orbiter
MSR = Mars Sample Return
MSRL = Mars Sample Retrieval Lander
MSRM = Mars Sample Return Mission
MSRV = Mars Sample Return Vehicle
NASA = National Aeronautics and Space Administration
OR = Orbital Rendezvous
SFR = Sample Fetching Rover

## HISTORY OF WORKS & PRESENTATIONS

This novel proposed concept was first orally presented at International 2020 AIAA-YPSE Student Conference (Mid-Atlantic section) held virtually from 15-16 October 2020 at United States. Then, the same proposal was orally presented at AIAA NextGen 2020 Technical Symposium held virtually from 22-23 October 2020 at Huntsville, United States. Later, full manuscript and presentation was submitted to 2021 AIAA/IEEE Aerospace Conference held virtually from 6-13 March 2021 at Yellowstone Conference Center, Big sky, Montana, United States.

## APPENDICES

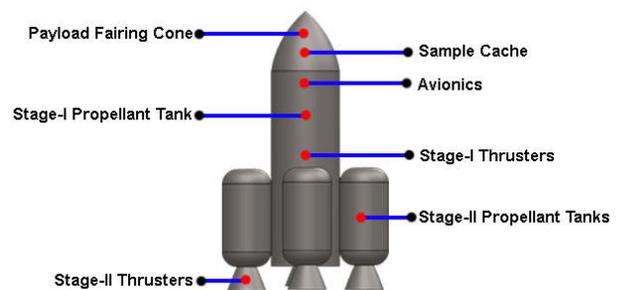

**TYPICAL DESIGN OF MARS SAMPLE RETURN VEHICLE**

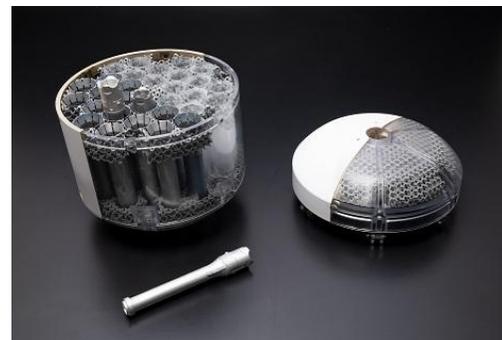

**Sample Cache Container of Mars 2020 Rover
(Image Courtesy: NASA)**

## BIOGRAPHY

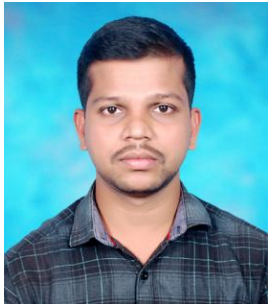

*Malaya Kumar Biswal M is a Bachelor of Science Graduate and a Graduate researcher at Department of Physics, Pondicherry University, India. His field of interest includes human exploration of Mars, planetary exploration, space case studies, and deep space transportation systems. He is actively involved in proposing mission strategies to enable Mars and Deep Space exploration. He is an active student member of the American Institute of Aeronautics and Astronautics (AIAA), USA, Indian Science Congress Association, India, and Mercury Exploration Assessment Group (MEXAG). Beginning his earlier research career since undergraduate level, he holds reputed publications from AIAA & IAA conferences, and the Mars Society.*
*Contact: malaykumar1997@gmail.com*

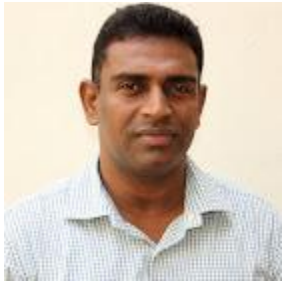

*Ramesh Naidu Annavarapu is a Doctorate in Physics from University of Hyderabad, India. He is currently working as an Associate Professor at the Department of Physics, Pondicherry University. He obtained N. V. Ranganayakamma medal at University of Hyderabad. He has thrice cleared the CSIR-UGC-NET exam. He has done his Post-Doctoral research at the Neuroscience Centre, University of Helsinki, Helsinki, Finland. His research areas include Theoretical Physics, Computational Physics and Computational Neuroscience. He is more interested in the interdisciplinary research areas where the fundamental laws of Physics can be applied. He has published two books and more than 20 research articles. He was a member of International Society for Gestalt Theory. He is an editorial and review board member of the journal "Computational and Applied Science".*
*Contact: rameshnaidu.phy@pondiuni.edu.in*